\documentclass[twocolumn]{revtex4-1}
\usepackage{amsmath,amssymb,graphicx}
\begin{document}

\title{Spatiotemporal Characterization of Supercontinuum Extending from the Visible to the Mid-Infrared in Multimode Graded-Index Optical Fiber}

\author{K. Krupa}
\affiliation{Universit\'e de Limoges, XLIM, UMR CNRS 7252, 123 Avenue A. Thomas, 87060 Limoges, France}
\affiliation{Universit\'e de Bourgogne Franche-Comt\'e, ICB UMR CNRS 6303, 9 Avenue A. Savary 21078 Dijon, France}
\author{C. Louot}
\affiliation{Universit\'e de Limoges, XLIM, UMR CNRS 7252, 123 Avenue A. Thomas, 87060 Limoges, France}
\author{V. Couderc}
\affiliation{Universit\'e de Limoges, XLIM, UMR CNRS 7252, 123 Avenue A. Thomas, 87060 Limoges, France}
\author{M. Fabert}
\affiliation{Universit\'e de Limoges, XLIM, UMR CNRS 7252, 123 Avenue A. Thomas, 87060 Limoges, France}
\author{R. Guenard}
\affiliation{Universit\'e de Limoges, XLIM, UMR CNRS 7252, 123 Avenue A. Thomas, 87060 Limoges, France}
\author{B. M. Shalaby}
\affiliation{Universit\'e de Limoges, XLIM, UMR CNRS 7252, 123 Avenue A. Thomas, 87060 Limoges, France}
\affiliation{Physics Department, Faculty of Science, Tanta University, 31527, Tanta, Egypt }
\author{A. Tonello}\email[]{Corresponding author: alessandro.tonello@unilim.fr}
\affiliation{Universit\'e de Limoges, XLIM, UMR CNRS 7252, 123 Avenue A. Thomas, 87060 Limoges, France}
\author{A. Barth\'el\'emy}
\affiliation{Universit\'e de Limoges, XLIM, UMR CNRS 7252, 123 Avenue A. Thomas, 87060 Limoges, France}
\author{D. Pagnoux}
\affiliation{Universit\'e de Limoges, XLIM, UMR CNRS 7252, 123 Avenue A. Thomas, 87060 Limoges, France}
\author{P. Leproux}
\affiliation{Universit\'e de Limoges, XLIM, UMR CNRS 7252, 123 Avenue A. Thomas, 87060 Limoges, France}
\author{A. Bendahmane}
\affiliation{Universit\'e de Bourgogne Franche-Comt\'e, ICB UMR CNRS 6303, 9 Avenue A. Savary 21078 Dijon, France}
\author{R. Dupiol}
\affiliation{Universit\'e de Bourgogne Franche-Comt\'e, ICB UMR CNRS 6303, 9 Avenue A. Savary 21078 Dijon, France}
\author{G. Millot}
\affiliation{Universit\'e de Bourgogne Franche-Comt\'e, ICB UMR CNRS 6303, 9 Avenue A. Savary 21078 Dijon, France}
\author{S. Wabnitz}
\affiliation{Dipartimento di Ingegneria dell'Informazione and INO-CNR, Universit\`a di Brescia, via Branze 38, 25123, Brescia, Italy }

\begin{abstract}
We experimentally demonstrate that pumping a graded-index multimode fiber with sub-ns pulses from a microchip Nd:YAG laser leads to spectrally flat supercontinuum generation with a uniform bell-shaped spatial beam profile extending from the visible to the mid-infrared at 2500\,nm. We study the development of the supercontinuum along the multimode fiber by the cut-back method, which permits us to analyze the competition between the Kerr-induced geometric parametric instability and stimulated Raman scattering. We also performed a spectrally resolved temporal analysis of the supercontinuum emission.
\end{abstract}

\maketitle

The strong modal confinement and the versatile dispersion engineering of single mode fibers (SMFs) have permitted to demonstrate efficient and spatially coherent supercontinuum (SC) sources spanning from the ultra-violet to the mid-infrared (MIR) \cite{dudleycoen}. However, SMF have an inherent major drawback, namely the relatively low energy (typically less than 
20\,$\mu$J for pulses slightly shorter than 1\,ns) that can be accepted owing to their small mode area. 
 
 For this reason, SC sources based on SMFs cannot be used for applications where high pulse energies are required, such as airborne remote sensing and beam delivery. On the other hand, it is well known that multimode fibers (MMFs) such as graded-index (GRIN) fibers, which support the propagation of 100-1000 guided modes, can permit the propagation of pulses of higher energy but are subject to strong mode-mixing. This fact results from the difference of propagation constants of the modes and from the
 linear mode coupling among nearly degenerate groups of modes, which spoils the initial spatial coherence of a light beam after a few millimeter of propagation. As a result of modal interference, a speckled intensity pattern results at the MMF output, which prevents the use of MMFs whenever the preservation of spatial coherence
 is required \cite{Picozzi2015R30}. 

However recent experiments by Krupa et al. \cite{KrupaClean} led to the unexpected discovery that the cubic, or Kerr, nonlinearity of glass fibers may lead, above a certain threshold pulse power, to the generation of a self-sustained spatially coherent nonlinear beam in a highly multimode GRIN fiber. This means that linear mode mixing can be effectively washed out by means of the Kerr effect, so that a cleaned multimode light beam remains effectively self-preserved. The mechanism for Kerr beam self-cleaning is the nonreciprocitity of the nonlinear coupling among the fundamental mode and the higher-order modes \cite{KrupaClean}, which is induced by the dynamic longitudinal grating resulting from periodic mode beating and the Kerr effect \cite{schnack15}. 
It is important to note that Kerr self-cleaning occurs at peak pulse power levels which are at least one order of magnitude lower than the critical power associated with the well-known effect of catastrophic light self-focusing in a GRIN MMF \cite{Mana88}, therefore the linear mode basis can still be considered for the expansion of the multimode field. Moreover, the Kerr-induced recovery of spatial coherence is a conservative process which occurs for the pump beam itself, and it essentially involves a lossless power transfer into the fundamental mode. Its nature is thus basically different from the well-known dissipative beam clean-up that is observed at the Stokes wavelength, resulting from modal-dependent Raman gain \cite{RamanClean}.
In addition, Kerr self-cleaning occurs before a substantial pump spectral broadening has occurred \cite{KrupaClean}.

For powers above the self-cleaning threshold, nonlinear spectral broadening in MMFs results from a complex interplay between spatial and temporal degrees of freedom \cite{WiseNatPhoton}. 
Because of the self-imaging of multimode beams in a GRIN MMF, the Kerr effect leads to a long-period intensity grating which induces mode conversion \cite{schnack15} and quasi-phase-matched (QPM) four-wave-mixing (FWM) \cite{LonghiOL}. For temporal multimode femtosecond solitons in the anomalous dispersion regime \cite{Renninger2012R31}, the nonlinear index grating produces an effective periodic nonlinearity, which in turn induces a series of dispersive wave sidebands  \cite{WiseNatPhoton,WisePRL}.  On the other hand, for a quasi-CW pump pulse propagating in the normal dispersion regime (in order to avoid the modulation instability induced break-up into an ultrashort soliton pulse train)  the nonlinear index grating leads to a series of intense QPM-FWM (or geometric parametric instability (GPI)) sidebands in the visible and near-infrared (NIR) \cite{KrupaPRL}.
Quite remarkably, experiments of Ref.\cite{KrupaPRL} revealed that the Kerr self-cleaning of the pump beam is frequency transferred across the entire spectrum of the GPI sidebands .
Finally, for higher powers and fiber lengths, the GPI sidebands merge into a broadband SC as a result of the interplay of the Kerr effect and Raman scattering, and both effects cooperate with each other in reinforcing the wideband spatial beam cleaning \cite{wright2016,Correa}.

In this Letter we carry out a spatio-temporal experimental characterization of SC generation in GRIN MMFs, extending from the visible to NIR and MIR with remarkable spectral flatness.  We focus our attention to point out
the respective role of the Kerr and Raman scattering mechanisms in determining both spectral broadening and beam cleaning. By cut-back measurements we can identify the respective length scale of development for the visible and NIR spectral components. By spectrally resolved temporal measurements, we 
can determine the temporal structures associated with the GPI sidebands, the depleted pump and the Stokes light at NIR wavelengths. 

\begin{figure}[htbp]
  \centering  
     \includegraphics[width=0.9\linewidth]{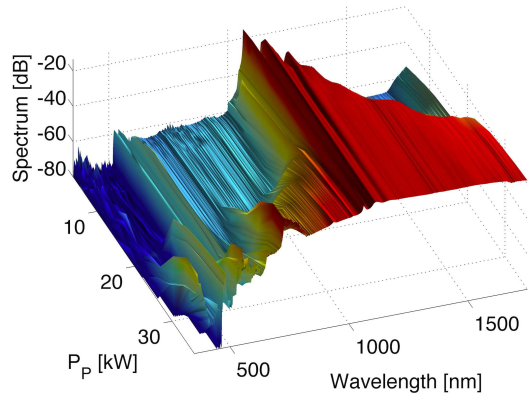}
    \caption{Experimental  SC spectra as a function of the pump pulse peak power $P_p$, 
    for a 12\,m long GRIN MMF.}
\label{fig:1}
\end{figure}

In our experiments, we used as pump source an amplified Nd:YAG microchip laser, delivering 
900\,ps pulses at 1064\,nm with 30\,kHz repetition rate. The
linearly polarized Gaussian pump pulses were launched into the GRIN MMF (core diameter 52\,$\mu m$, NA=0.2) 
by using a lens with focal length of 
$50\,mm$ and a three-axis translating stage.
 At the input face of the fiber the beam had a FWHMI diameter of $40\,\mu m$,
which is close to the fiber core diameter.

\begin{figure}[htbp]
  \centering  
    \includegraphics[width=0.9\linewidth]{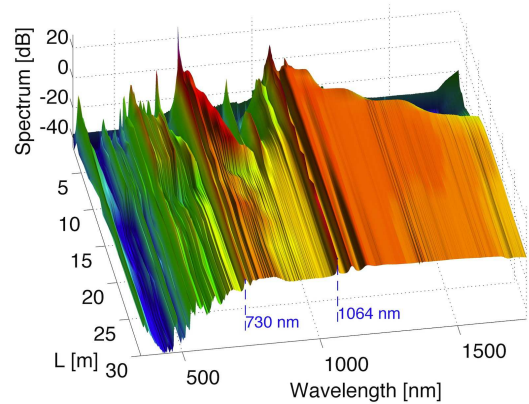}
\caption{Experimental results of the cut-back technique (input average power in the fiber 1.4\,W, drops to 670\,mW after 30\,m of fiber). All experimental spectra have been visualized by normalizing to the level at 1118~nm.}
\label{fig:2}
\end{figure}

We used an optical spectrum analyzer (OSA) covering the spectral range from 350\,nm up to 1750\,nm,
as well as a spectrometer covering the range from 850\,nm till 2500\,nm. The   
beam profile (at the output face of the fiber) was imaged on a CCD camera through a microlens with 8\,mm focal length. 
 Different 10\,nm-wide bandpass interference optical filters were introduced in order to
characterize the spatial beam profile at the pump wavelength, and at the other wavelengths across the entire SC. Note that, owing to the long pump pulse
duration and to the low modal GRIN MMF dispersion, a large number of initially excited modes maintain their temporal superposition, hence their
nonlinear coupling, over tens of meters.

\begin{figure}[t]
  \centering 
    \includegraphics[width=0.85\linewidth]{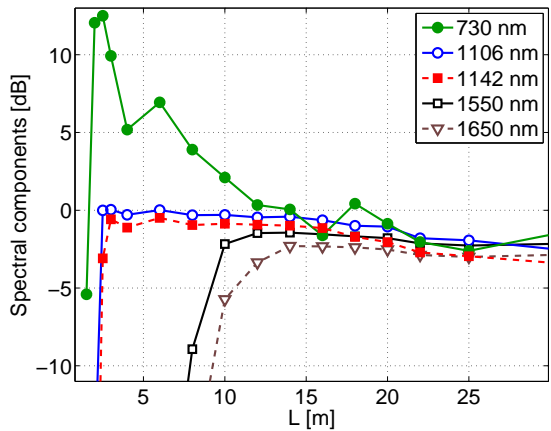}
    \caption{Detail of the spatial growth of intensity at selected spectral components from Fig. \ref{fig:2}. The spectral components are normalized to the level at 1118\,nm}
\label{fig:3}
\end{figure}

Fig. \ref{fig:1}  provides the experimental characterization of the optical spectrum at the output of a 12\,m long
GRIN MMF, as a function of the peak power of the pump pulses. 
As can be seen, a visible SC develops until right below 500\,nm whenever the pump peak power grows larger than 25\,kW.
In terms of spectral flatness, as the peak power reaches 35\,kW the SC amplitude remains relatively uniform until about 700\,nm, beyond which it rapidly drops. 
On the other hand, spectral broadening on the infrared (IR) side of the pump continuously grows with pump power: above 25\,kW of peak power, the SC extends beyond 1750\,nm with remarkable spectral flatness.
For understanding the dynamics of SC generation in a GRIN MMF under the combined action of Kerr effect and Raman scattering, it is crucial to study the growth of SC along the fiber length. Fig. \ref{fig:2} displays the results of cut-back experiments that describe the development of SC along the MMF, for a fixed input average power of 1.4\,W.

\begin{figure}[t]
 \centering
    \includegraphics[width=0.9\linewidth]{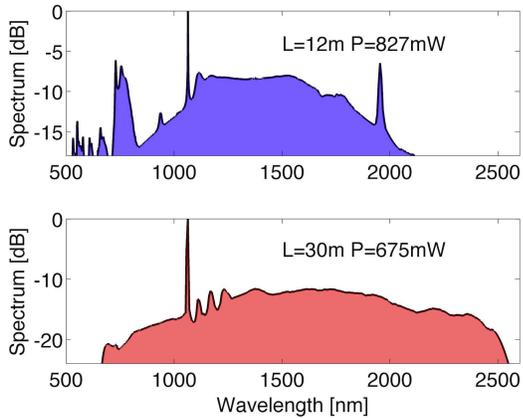}
    \caption{Comparison of supercontinuum intensity profiles 12\,m (top) and 30\,m (bottom) of GRIN MMF}
\label{fig:5}
\end{figure}

As can be seen GPI and Raman provide the major contributions to the spectral broadening, but their effects come with different scales. In the very first meter of propagation, GPI generates the first-order anti-Stokes sidebands at about 730\,nm which carries around 12\% of the total input power \cite{LonghiOL,KrupaPRL}. After 1.5\,m of propagation the Raman effect comes into play causing a gradual spectral extension toward the infrared side of the spectrum: the resulting pump depletion in turn inhibits the GPI frequency conversion. For fiber lengths till about 12\,m and for peak powers above 35\,kW the Raman induced SC spectrum assumes an unusual flat shape in the IR (see also later the upper panel of figure \ref{fig:5}), where one would expect instead  sharp separated Raman peaks as it is often the case  for longer fiber  \cite{AgrawalGRIN}. In parallel, after the first 3\,m, the sideband at 730\,nm, which is no longer fed, undergoes Raman and cross-phase modulation spectral broadening so that the  gap between 730\,nm and 1000\,nm is gradually filled. The first three Raman Stokes sidebands between 1100\,nm and 1350\,nm can be identified for fiber lengths greater than 20\,m  (see also later the lower panel of Fig. \ref{fig:5}).

\begin{figure}[t]
  \centering 
  \includegraphics[width=0.85\linewidth]{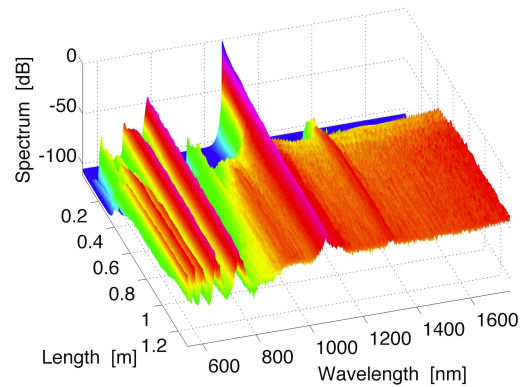}
  \caption{Numerical simulation of SC generation in GRIN MMF.}
\label{fig:4}
\end{figure}

More details about the longitudinal development of SC can be extracted from Fig. \ref{fig:3}, where we compare the spatial evolution along the GRIN MMF of the power of selected spectral components extracted from the spectra of Fig. \ref{fig:2}. Fig. \ref{fig:3} clearly shows that components of NIR spectrum in between the Raman peaks  (components at 1106\,nm and at 1142\,nm) reach their maximum values after about 2\,m and have later on a gradual drop, consequently revealing the Raman sidebands. On the other hand the SC components in the telecom band (components at 1550\,nm and at 1650\,nm) only start to appear between 5\,m and 10\,m, and reach their stationary power levels above 15\,m of MMF.
Fig. \ref{fig:3} also reveals that the first anti-Stokes GPI sideband at 730\,nm initially overshoots to its peak power value within the first meter of fiber, and subsequently relaxes down to the same steady-state power level (situated about 15\,dB below the peak value) as all other components after 15\,m of fiber length.

We performed extensive numerical simulations to model the observed SC generation dynamics and for understanding the relative balance of the Kerr and Raman scattering mechanisms as a function of the MMF length. We have solved the nonlinear Schr\"odinger equation with two transverse dimensions and one temporal dimension, including a truncated parabolic profile for describing the GRIN MMF refractive index, with both Kerr nonlinearity and stimulated Raman scattering \cite{KrupaClean}.
In order to reduce the computational load, in our simulations we considered relatively short (below 1.2\,m) fiber lengths, by correspondingly increasing the pump pulse peak power and we considered a pump pulse duration of 9\,ps. 
Although this procedure is expected to only provide a qualitative correspondence with the experimental results, the numerical results of Fig. \ref{fig:4} fairly well predict the occurrence of sharp spectral peaks in the visible domain, as well as the development of a spectrally flat SC across the entire NIR. 
Note that in the spectra of Fig. \ref{fig:4} the transverse dimension has been integrated; moreover, the maximum intensity was of 10\,$GW/cm^2$ and the simulated fiber length was of 1.5\,m.

As shown in Fig. \ref{fig:5}, where we combined the output of the OSA and of the spectrometer, we observed that, as the MMF length increases up to 30\,m, SC generation can be extended into the MIR (i.e., up to 2500\,nm). The result of Fig. \ref{fig:5} is remarkable, when considering the large linear absorption of the MMF in the MIR. The comparison of the SC intensity profiles obtained for 12\,m and 30\,m reported in Fig. \ref{fig:5} clearly shows that the SC bandwidth critically depends upon the MMF length: hence its value should be properly optimized. Moreover, Fig. \ref{fig:5} reveals that the isolated GPI sidebands in the visible domain are better discernible for MMF lengths of about 10\,m (or shorter, as discussed with reference to Fig. \ref{fig:3}). On the other hand, spectral broadening is dominated by Raman scattering for MMF lengths above a few tens of meters, in agreement with earlier observations \cite{AgrawalGRIN}.

\begin{figure}[t]
  \centering
      \includegraphics[width=0.78\linewidth]{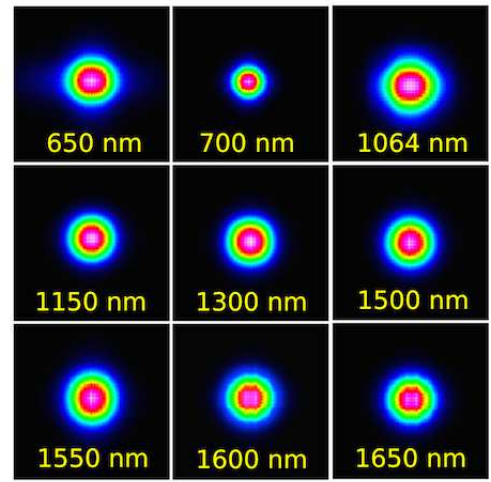}
      \caption{Far field at the output for L=30\,m, under different bandpass filters (10~nm bandwidth)and 1.4\,W input average power. }
\label{fig:7}
\end{figure}

Next we carried out a study of the wavelength dependence of the transverse beam profile emerging from the GRIN MMF. The plots of Fig. \ref{fig:7} show the far field at the output of 
the MMF analyzed by using a series of bandpass filters (10~nm bandwidth) with different center wavelengths, we observed that the spatial coherence acquired by the pump beam because of the Kerr self-cleaning effect \cite{KrupaClean} is effectively transferred across the entire spectrum of the SC. Fig. \ref{fig:7} shows that for all wavelengths a bell-shaped spatial profile is observed at the output of a GRIN MMF of length L=30\,m.  As we have discussed with reference to Fig. \ref{fig:5}, for this length of fiber the main mechanism for generating the NIR side of the SC is the Raman effect, which also privileges bell-shaped beams. However similar results for near and far fields (here not shown)   
were also observed in the visible domain as well as at the pump wavelength for fiber lengths shorter than 8\,m, that is well before the Raman scattering threshold  \cite{KrupaPRL}. 

\begin{figure}[t]
 \centering 
      \includegraphics[width=0.95\linewidth]{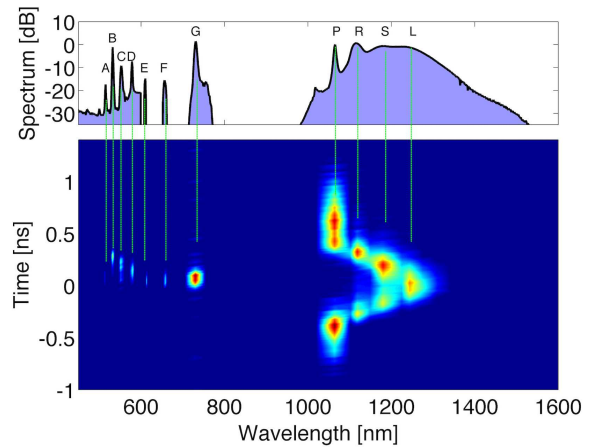}
\caption{Spectrum (upper panel) and Spectro-temporal characterization (lower panel)
of SC at the output of a 6\,m long GRIN MMF. Letters indicate the following wavelengths: 
515\,nm (A), 532\,nm (B), 552\,nm(C), 578\,nm(D), 610\,nm(E), 656\,nm (F), 730\,nm (G), 1064\,nm(P) 1116\,nm(R),
1180\,nm (L), 1250\,nm(S).}
\label{fig:6}
\end{figure}
  
For potential applications of MMF SC sources, such as in multiplex CARS spectroscopy \cite{multiVincent}, it is important to consider the possibility of achieving simultaneous interactions among multiple wavelengths on a sample. To clarify the spectro-temporal structure of the generated SC radiation, we experimentally analyzed the variation of the temporal envelope across different wavelengths, as shown in Fig. \ref{fig:6}. This was achieved by dispersing the fiber output light with a grating,
which permitted us to analyze the temporal envelope with two different 12-GHz photodiodes (in order to cover both the visible and the NIR portions of the SC) and a 20-GHz oscilloscope.
The limitations imposed by the electrical bandwidth of our detection system could not permit us to unveil fine spectro-temporal features below 83\,ps range; the intensity levels of the electric signals are in linear scale; intensities have been rescaled in proportion to the power spectrum carried by each optical spectral bandwidth. 
Nevertheless as shown in Fig. \ref{fig:6} our spectrogram revealed 
the presence of a deep temporal modulation within the flat portion of the NIR SC spectrum. 
In fact, Fig. \ref{fig:6} shows that the larger Raman frequency shift comes from the highest input peak power, 
which in turn leads to the strongest pump pulse depletion at the MMF output. As a consequence, the different power values that constitute the temporal profile of the input pump pulse are transposed into different wavelength shifts. This effect has already been experimentally characterized for SMFs in Ref.\cite{chirstophespie}.
Our observations confirm that a similar feature is also present in the long wavelength side of the pump for a GRIN MMF.  The relatively short (6\,m) MMF length could also permit us to clearly observe in Fig. \ref{fig:6} the temporal envelopes of the separate GPI sidebands. As can be seen, the GPI sideband durations are much shorter than the input pump pulse. A direct measurement of their  temporal duration is certainly limited by the electric bandwidth of our photodiode. Nevertheless the sidebands at 730\,nm, whose envelope covers 130\,ps can be directly revealed confirming the presence of complex space-time interplay and pulse shortening even by using nanosecond pump pulses.

In short summary, we have shown that multimode GRIN fibers provide a new, intriguing, low-cost and high-pulse-energy platform for SC generation ranging from the visible until the MIR. In this work we demonstrated that pumping a GRIN MMF with a sub-nanosecond microchip laser in the normal dispersion regime permits to maintain the spatial coherence of the SC, thanks to the cooperative action of lossless Kerr self-cleaning and dissipative Raman beam cleaning effects. We performed an experimental characterization of the growth rate of parametric and Raman components, and we described the time-frequency structure of the generated supercontinuum.

We acknowledge the financial support of 
Bpifrance OSEO and Horiba medical (Industrial Strategic Innovation Programme) with project Dat@diag.

\bibliography{refsMMF}

\end{document}